\RequirePackage{snapshot}

\documentclass[a4paper,twoside]{cernrep}

\makeatletter
\let\NAT@parse\undefined
\makeatother

\usepackage[numbers,sort,sectionbib]{natbib}
\usepackage[sectionbib]{bibunits}

\usepackage{times} 
\usepackage[T1]{fontenc} 
\usepackage[latin9]{inputenc} 
\usepackage{epic,eepic,pspicture,mathptmx,graphpap,epsf,titleref}

\usepackage{morefloats}
\usepackage{upgreek}
\usepackage{amsbsy,amsmath,amssymb} 
\usepackage{color,ulem} 
\usepackage{subfig}                 
\usepackage[printonlyused]{acronym} 
\usepackage[toc,page]{appendix}     
\usepackage{svn}                    
\usepackage{longtable}              
\usepackage{enumerate}		    
\usepackage{eurosym} 
\usepackage{multirow} 
\usepackage[loose]{units} 
\usepackage{threeparttable} 
\usepackage[point]{rccol} 
\usepackage{booktabs}
\usepackage{colortbl}
\usepackage{fancyhdr}
\usepackage{lineno}  
\usepackage{textpos} 
\usepackage{enumitem}
\setlist{noitemsep}
\setlist[itemize]{leftmargin=*, topsep=-0.3cm, partopsep=0pt, itemsep=0pt}
\setlist[enumerate]{leftmargin=*, topsep=0pt, partopsep=0pt, itemsep=0pt}

\usepackage[
     pdftex,
     plainpages        = false,
     bookmarks         = true,
     bookmarksnumbered = true,
     bookmarksopen     = true,
     pdfstartview      = FitH,
     pdfpagelayout     = SinglePage,
     colorlinks        = true,
     citecolor         = blue,
     linkcolor         = blue,
     urlcolor          = blue, 
     pdftitle={CLIC e+e- Linear Collider Studies},
     pdfauthor={CLIC},
     pdfkeywords={CLIC, accelerator, physics performances, detector concepts},
     pdfcreator={latex},
     pdfproducer={pdfLaTex}]{hyperref}

\def \clicsid {CLIC\_SiD\xspace}
\def \clicild {CLIC\_ILD\xspace}

\def\roots {\ensuremath{\sqrt{s}}\xspace}

\newcommand{\abinv}{\ensuremath{\mathrm{ab}^{-1}}\xspace}
\newcommand{\fbinv}{\ensuremath{\mathrm{fb}^{-1}}\xspace}

\def\epem       {\ensuremath{e^+e^-}\xspace}

\newcommand{\modeltwo}{\textit{model~II}\xspace}
\newcommand{\modelthree}{\textit{model~III}\xspace}
\def\sqsq #1 {\ensuremath{\PSQ_{\mathrm{#1}}\PSQ_{\mathrm{#1}}}\xspace} 

\newcommand{\micron}{\ensuremath{\upmu\mathrm{m}}\xspace}

\begin{document}

 \begin{textblock*}{0mm}(135mm,-18mm)%
 \includegraphics[width=3cm]{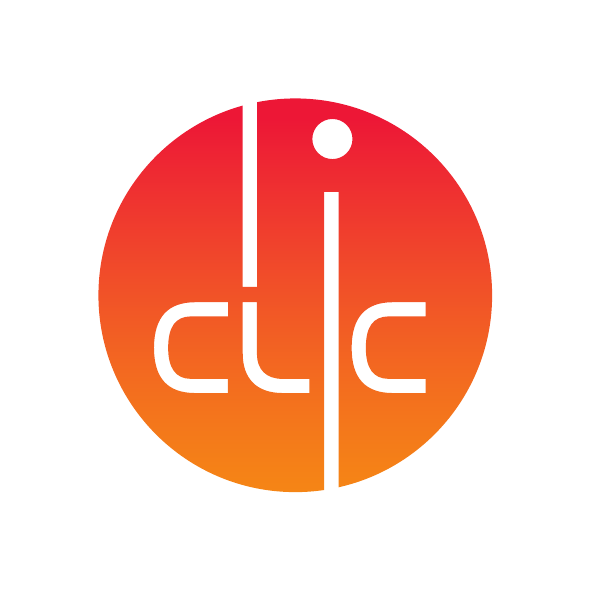}%
 \end{textblock*}%
\vspace{-2cm}
\begin{center}
{\huge  {CLIC} \epem Linear Collider Studies}\\\vspace{5mm}
{\LARGE  Input to the update process of the\\}
{\LARGE  European Strategy for Particle Physics}
\end{center}


\begin{center}
July $31^{\mathrm{st}}$, 2012
\end{center}

\begin{center}
This document provides input from the CLIC \epem linear collider studies to the update process of the European Strategy for Particle Physics. It is submitted on behalf of the CLIC/CTF3 collaboration and the CLIC physics and detector study.
\end{center}

\begin{center}
Corresponding editors: Dominik Dannheim, Philippe Lebrun, Lucie
Linssen, Daniel Schulte,\\
 Frank Simon, Steinar Stapnes, Nobukazu Toge, Harry Weerts, James Wells
\end{center}

\section{Introduction}

The Compact Linear Collider (CLIC) is a TeV-scale high-luminosity
linear \epem collider under development. It is based on a novel
two-beam acceleration technique providing acceleration gradients at
the level of 100 MV/m.  A high-luminosity high-energy \epem collider
allows for the exploration of Standard Model (SM) physics, such as precise
measurements of the Higgs, top and gauge sectors, as well as for a
multitude of searches for New Physics, either through direct discovery
or indirectly, via high-precision observables.  Given the current
state of knowledge, following the observation of a $\sim$125 GeV
Higgs-like particle at the LHC, and pending further LHC results at 8
TeV and 14 TeV, a linear \epem collider built and operated in
centre-of-mass energy stages from a few-hundred GeV up to a few TeV
will be an ideal physics exploration tool, complementing the LHC.
 
This document provides short summaries of the CLIC accelerator design,
performances and implementation studies, the layout and performances
of the CLIC experiments under study and the projected CLIC physics
potential, followed by an outlook on the CLIC programme in the coming
years. For more detailed descriptions we refer to the following
documents: \vspace{2mm}
\begin{itemize}
\item The Physics Case for an \epem Linear Collider, eds. J. Brau et
  al., submitted to the update process of the European Strategy for
  Particle Physics, July 2012~\cite{LC_Phys_Case_Strategy_Input};
\item A Multi-TeV Linear Collider based on CLIC Technology, CLIC
  Conceptual Design Report, 2012, eds.  M. Aicheler et
  al.~\cite{CLICCDR_vol1};
\item Physics and Detectors at CLIC, CLIC Conceptual Design Report,
  eds. L. Linssen et al.~\cite{CLICCDR_vol2};
\item The CLIC Programme: towards a staged \epem Linear Collider
  exploring the Terascale, CLIC Conceptual Design Report, 2012,
  eds. P. Lebrun et al.~\cite{CLICCDR_vol3}.
\end{itemize}
\vspace{2mm} The above CLIC CDR reports are supported by more than
1300
signatories\footnote{\href{https://edms.cern.ch/document/1183227/}{https://edms.cern.ch/document/1183227/}}
from the world-wide particle physics community.

\section{Accelerator Complex}

\begin{figure}[b!]
  \centering
  \includegraphics[scale=0.82]{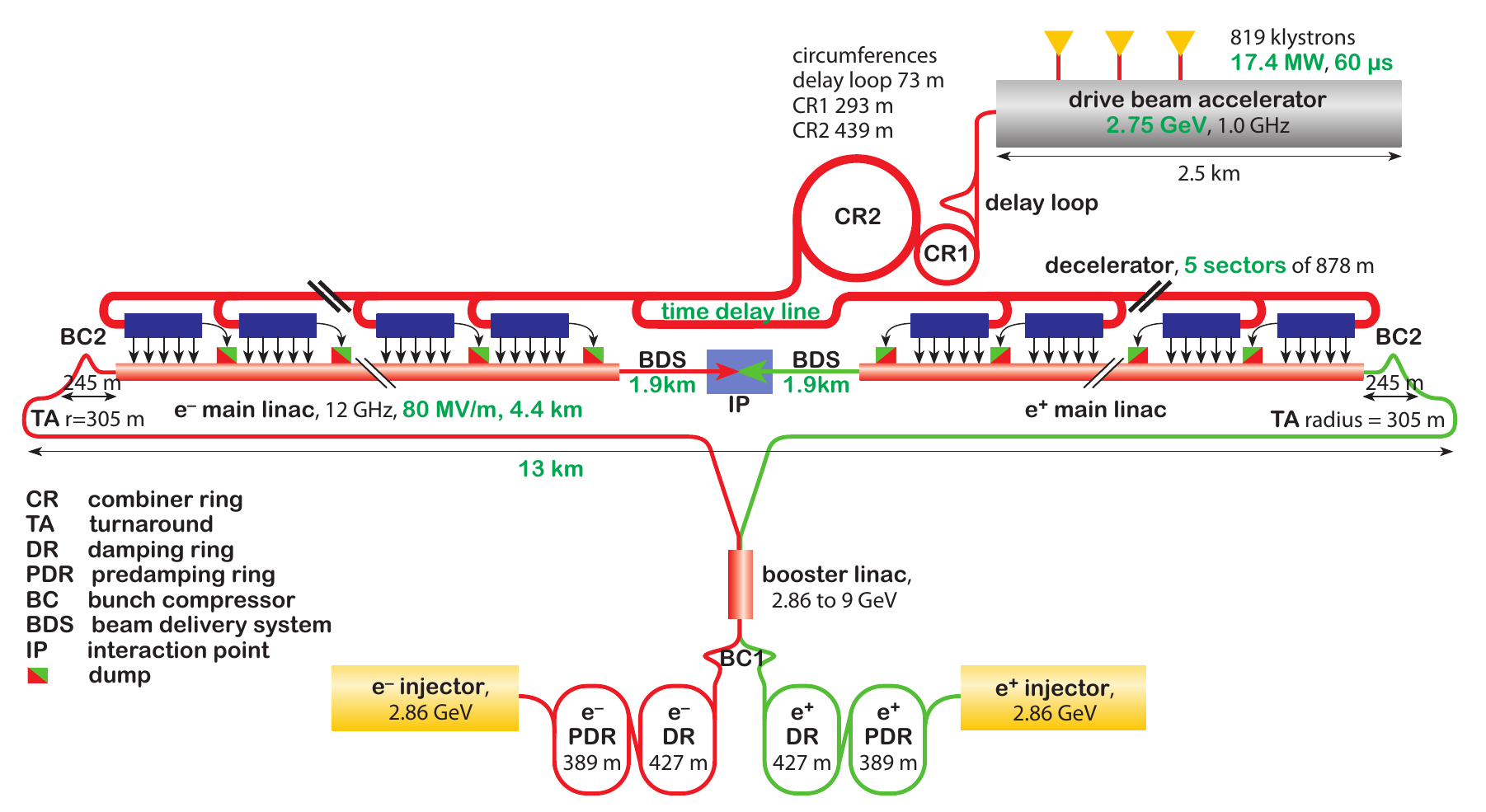}
  \caption{Overview of the CLIC layout at $\sqrt{s}=500$~GeV (scenario
    A).}
  \label{f:clic500gev}
\end{figure}

\begin{figure}[b!]
  \centering
  \includegraphics[scale=0.82]{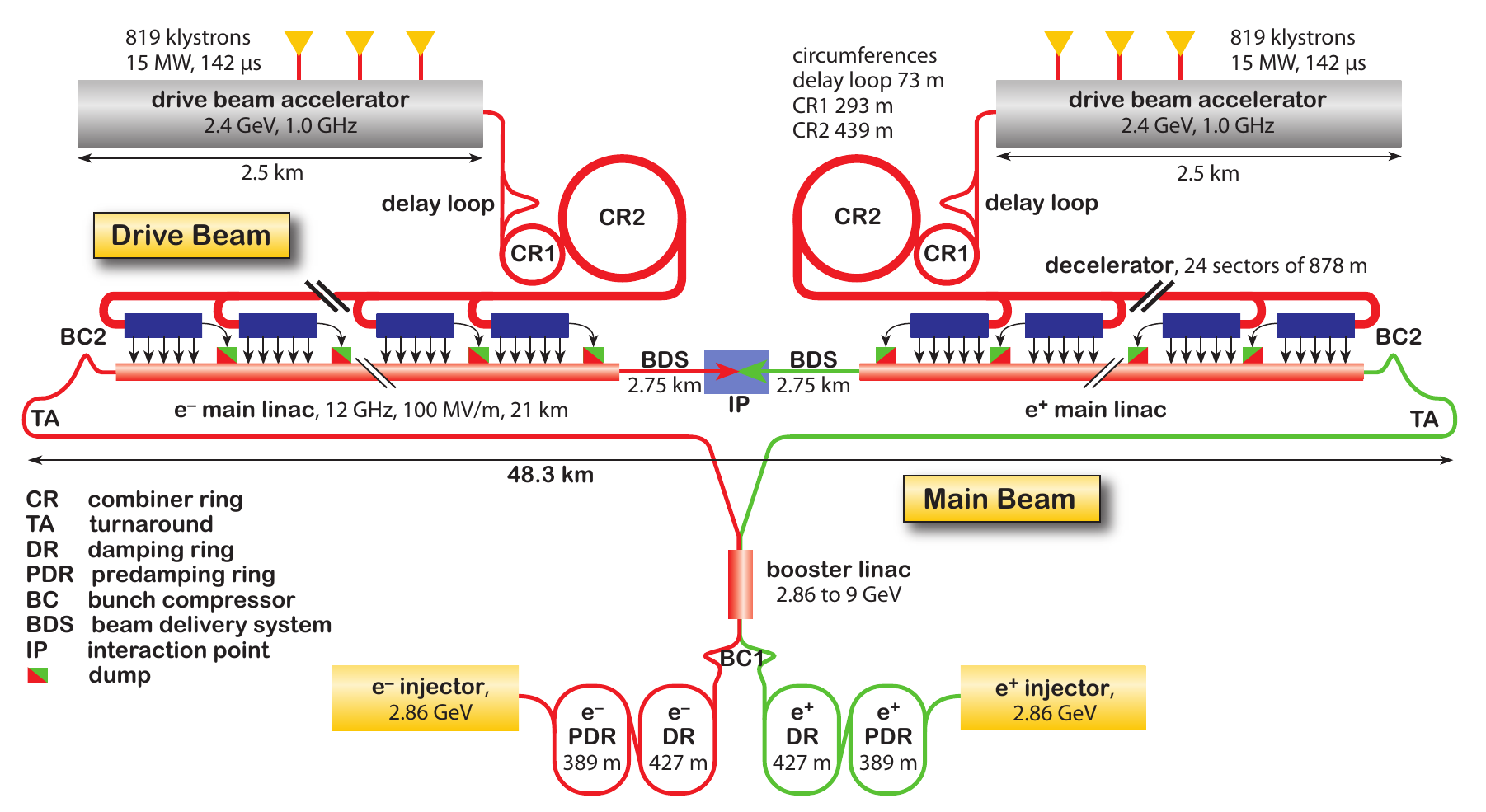}
  \caption{Overview of the CLIC layout at $\sqrt{s}=3$~TeV.}
  \label{f:clic3tev}
\end{figure}

The CLIC layouts at 500~GeV and 3~TeV are shown in
Figures~\ref{f:clic500gev} and \ref{f:clic3tev}, respectively, and the
key parameters are given in Tables~\ref{t:1} and \ref{t:2}.  The
conceptual design is detailed in \cite{CLICCDR_vol1} and
\cite{CLICCDR_vol3}.
The CLIC design is based on three key technologies, which have been
addressed experimentally:\vspace{2mm}
\begin{itemize}
\item The use of normal-conducting accelerating structures in the main
  linac with a gradient of $100\;\rm MV/m$, in order to limit the
  length of the machine.  The RF frequency of $12\;\rm GHz$ and
  detailed parameters of the structure have been derived from an
  overall cost optimisation at 3~TeV. Experiments at KEK, SLAC and
  CERN verified the structure design and established its gradient and
  breakdown-rate performance.
\item The use of drive beams that run parallel to the colliding beams
  through a sequence of power extraction and transfer structures,
  where they produce the short, high-power RF pulses that are
  transferred into the accelerating structures. These drive beams are
  generated in a central complex.  The drive-beam generation and use
  has been demonstrated in a dedicated test facility (CTF3) that has
  been constructed and operated for many years at CERN by the
  CLIC/CTF3 collaboration.
\item The high luminosity that is achieved by the very small beam
  emittances, which are generated in the damping rings and maintained
  during the transport to the collision point. These emittances are
  ensured by appropriate design of the beam lines and tuning
  techniques, as well as by a precision pre-alignment system and an
  active stabilisation system that decouples the magnets from the
  ground motion. Prototypes of both systems have demonstrated
  performance close to or better than the specifications.
\end{itemize} \vspace{2mm} 
Related system parameters have been benchmarked in CTF3, in advanced light
sources, ATF(2) and CesrTA, and in other setups. 
In addition, a broad technical development programme has successfully
addressed many critical components. Among them are those of the main
linac, which are most important for the cost, and their integration
into modules. The drive-beam components have largely been addressed in
CTF3. Other performance-critical components have been developed and
tested, e.g., the final focus magnets, which will be located in the
detector and need to provide a very high field, and high-field damping
ring wigglers, which rapidly reduce the beam emittances.  Design
studies foresee 80\% polarisation of the electrons at collision, and
the layout is compatible with addition of a polarised positron source.
The successful validation of the key technologies and of the critical
components establish confidence that the CLIC performance goals can be
met.

Several of these technologies have applications for and are being
developed with other communities, e.g., synchrotron light sources, free
electron lasers and medical accelerators.

\begin{figure}[tb]
  \centering
  \includegraphics[width=130mm]{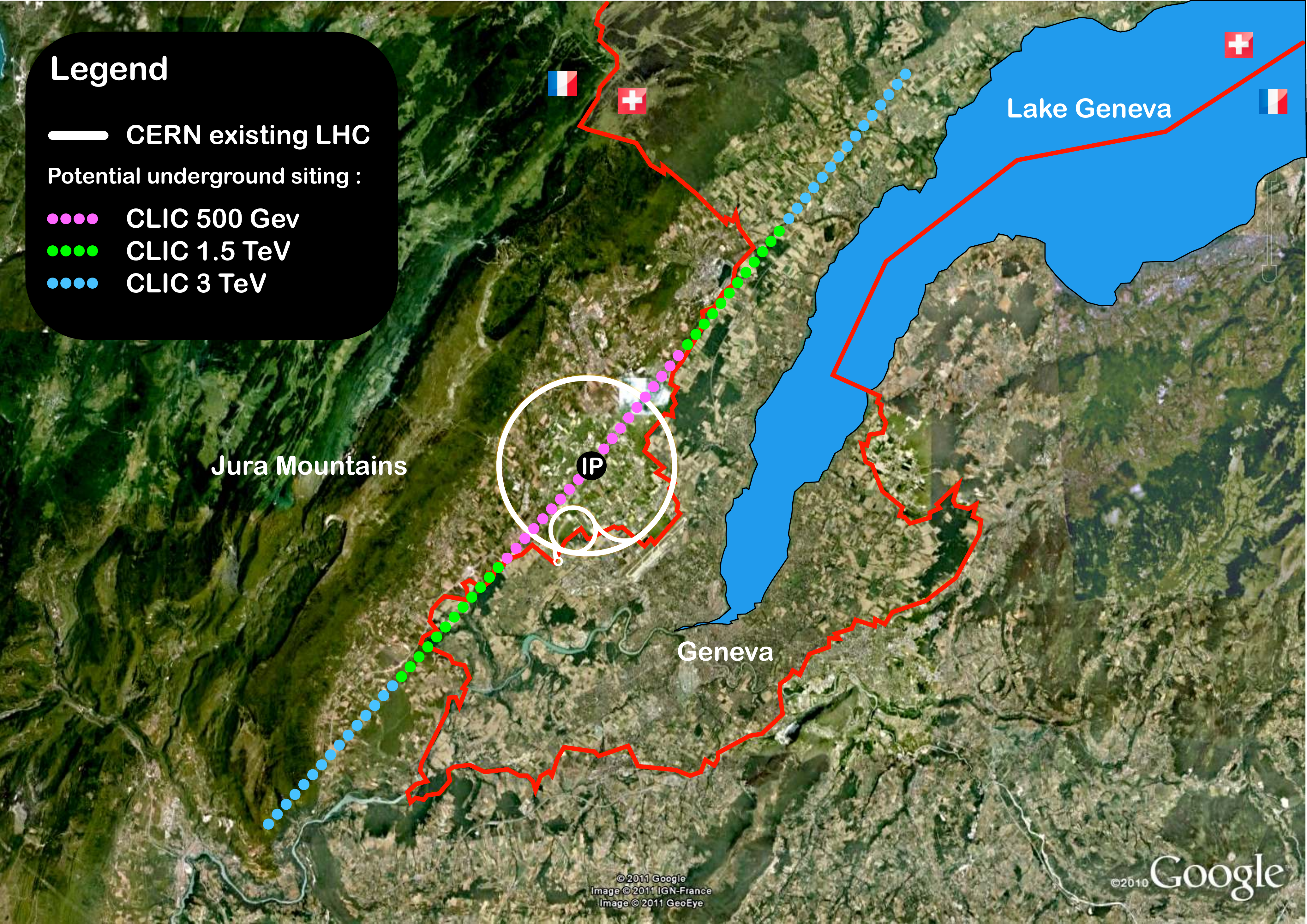}
  \caption{CLIC footprints near CERN, showing various implementation
    stages.}
  \label{fig:site}
\end{figure}

Detailed site studies show that CLIC can be implemented underground
near CERN, with the central main and drive beam complex on the CERN
domain, as shown in Figure~\ref{fig:site}. The site specifications do
not constrain the implementation to this location.

As indicated above, the current CLIC parameters are the result of a cost optimisation at
$3\;\rm TeV$, see Chapter 2.1 in~\cite{CLICCDR_vol1}.  However, the technology
can be used effectively over a wide range of centre-of-mass
energies. The project can be built in energy stages, which can re-use
the existing equipment for each new stage.  At each energy stage the
centre-of-mass energy can be tuned to lower values within a range of a
factor three and with limited loss on luminosity performance.  Two
example scenarios of energy staging are given in~\cite{CLICCDR_vol3}
with stages of $500\;\rm GeV$, $1.4~(1.5)\;\rm TeV$ and $3\;\rm TeV$,
see Table~\ref{t:1} for scenario A and Table~\ref{t:2} for scenario B.
For both scenarios the first and second stage use only a single
drive-beam generation complex to feed both linacs, while in stage 3
each linac is fed by a separate complex. Based on future physics
findings, the choice of energy stages will be reviewed and the design
optimised.  In case of growing interest in a lower energy Higgs
factory, studies of a klystron-based initial stage with a faster
implementation could become part of this evaluation.
\begin{table}[ht!]
  \centering
  \caption{Parameters for the CLIC energy stages of scenario A.}
  \label{t:1}
  \begin{tabular}{l l l l l l}
    \toprule
    \textbf{Parameter} & \textbf{Symbol} & \textbf{Unit}& \textbf{Stage 1} & \textbf{Stage
      2} & \textbf{Stage 3}\\
    \midrule
    Centre-of-mass energy            & $\roots$              &GeV                                            & 500 & 1400 & 3000\\
    Repetition frequency             & $f_{rep}$             &Hz                                             & 50 & 50 & 50\\
    Number of bunches per train      & $n_{b}$               &                                               & 354 & 312 & 312\\
    Bunch separation                 & $\Delta\,t$           &ns                                             & 0.5 & 0.5 & 0.5\\
    \midrule
    Accelerating gradient       & $G$                        &MV/m                                           & 80 & 80/100 & 100\\
    \midrule
    Total luminosity                 & $\mathcal{L}$         &$10^{34}\;\textrm{cm}^{-2}\textrm{s}^{-1}$  & 2.3 & 3.2 & 5.9 \\
    Luminosity above 99\% of $\roots$& $\mathcal{L}_{0.01}$  &$10^{34}\;\textrm{cm}^{-2}\textrm{s}^{-1}$  & 1.4 & 1.3 & 2\\
    \midrule
    Main tunnel length          &                            &km                                             & 13.2 & 27.2 & 48.3\\
    Charge per bunch            & $N$                        &$10^9$                                         & 6.8 & 3.7 & 3.7\\
    Bunch length                & $\sigma_z$                 &$\micron$                                      & 72 & 44 & 44\\
    IP beam size                & $\sigma_x/\sigma_y$        &nm                                             & 200/2.6 & $\sim$ 60/1.5 &$\sim$ 40/1\\
    Normalised emittance (end of linac) & $\epsilon_x/\epsilon_y$    &nm                                             & 2350/20 & 660/20 & 660/20\\
    Normalised emittance (IP)        & $\epsilon_x/\epsilon_y$    &nm                                             & 2400/25 & --- & ---\\
    Estimated power consumption & $P_{wall}$                 &MW                                             & 272 & 364 & 589\\
    \bottomrule
  \end{tabular}
\end{table}
\begin{table}[ht!]
  \caption{Parameters for the CLIC energy stages of scenario B.}
  \label{t:2}
  \centering
  \begin{tabular}{l l l l l l}
    \toprule
    \textbf{Parameter}           & \textbf{Symbol} & \textbf{Unit}& \textbf{Stage 1} & \textbf{Stage
      2} & \textbf{Stage 3} \\
    \midrule
    Centre-of-mass energy             & $\roots$                &GeV                                           & 500 & 1500 & 3000\\
    Repetition frequency              & $f_{rep}$               &Hz                                            & 50 & 50 & 50\\
    Number of bunches per train       & $n_{b}$                 &                                              & 312 & 312 & 312\\
    Bunch separation                  & $\Delta\,t$            &ns                                            & 0.5 & 0.5 & 0.5\\
    \midrule
    Accelerating gradient                     & $G$                     &MV/m                                          & 100 & 100 & 100\\
    \midrule
    Total luminosity                  & $\mathcal{L}$           &$10^{34}\;\textrm{cm}^{-2}\textrm{s}^{-1}$ & 1.3 & 3.7 & 5.9 \\
    Luminosity above 99\% of $\roots$ & $\mathcal{L}_{0.01}$    &$10^{34}\;\textrm{cm}^{-2}\textrm{s}^{-1}$ & 0.7 & 1.4 & 2\\
    \midrule
    Main tunnel length                &                         &km                                            & 11.4 & 27.2 & 48.3\\
    Charge per bunch                  & $N$                     &$10^9$                                      & 3.7 & 3.7 & 3.7\\
    Bunch length                      & $\sigma_z$              &$\micron$                                    & 44 & 44 & 44\\
    IP beam size                      & $\sigma_x/\sigma_y$     &nm                                            & 100/2.6 & $\sim$ 60/1.5 & $\sim$ 40/1\\
    Normalised emittance (end of linac)    & $\epsilon_x/\epsilon_y$ &nm                                            & --- & 660/20 & 660/20\\
    Normalised emittance                   & $\epsilon_x/\epsilon_y$ &nm                                            & 660/25 & --- &---\\
    Estimated power consumption       & $P_{wall}$              &MW                                            & 235 & 364 & 589\\
    \bottomrule
  \end{tabular}
\end{table}

Staging scenario A aims at achieving high luminosity at 500~GeV
collision energy with increased beam current. This requires
larger apertures in the accelerating structures which
therefore operate at a lower gradient. The re-use of these structures
in the second stage limits the achievable collision energy to
1.4~TeV. Staging scenario B aims at reducing the cost of the 500~GeV
stage using full-gradient accelerating structures at nominal beam
current, resulting in lower instantaneous luminosity.  The re-use of
these structures allows reaching 1.5~TeV collision energy in the
second stage.

\begin{figure}[t!]
  \centering
  \includegraphics[width=0.45\textwidth]{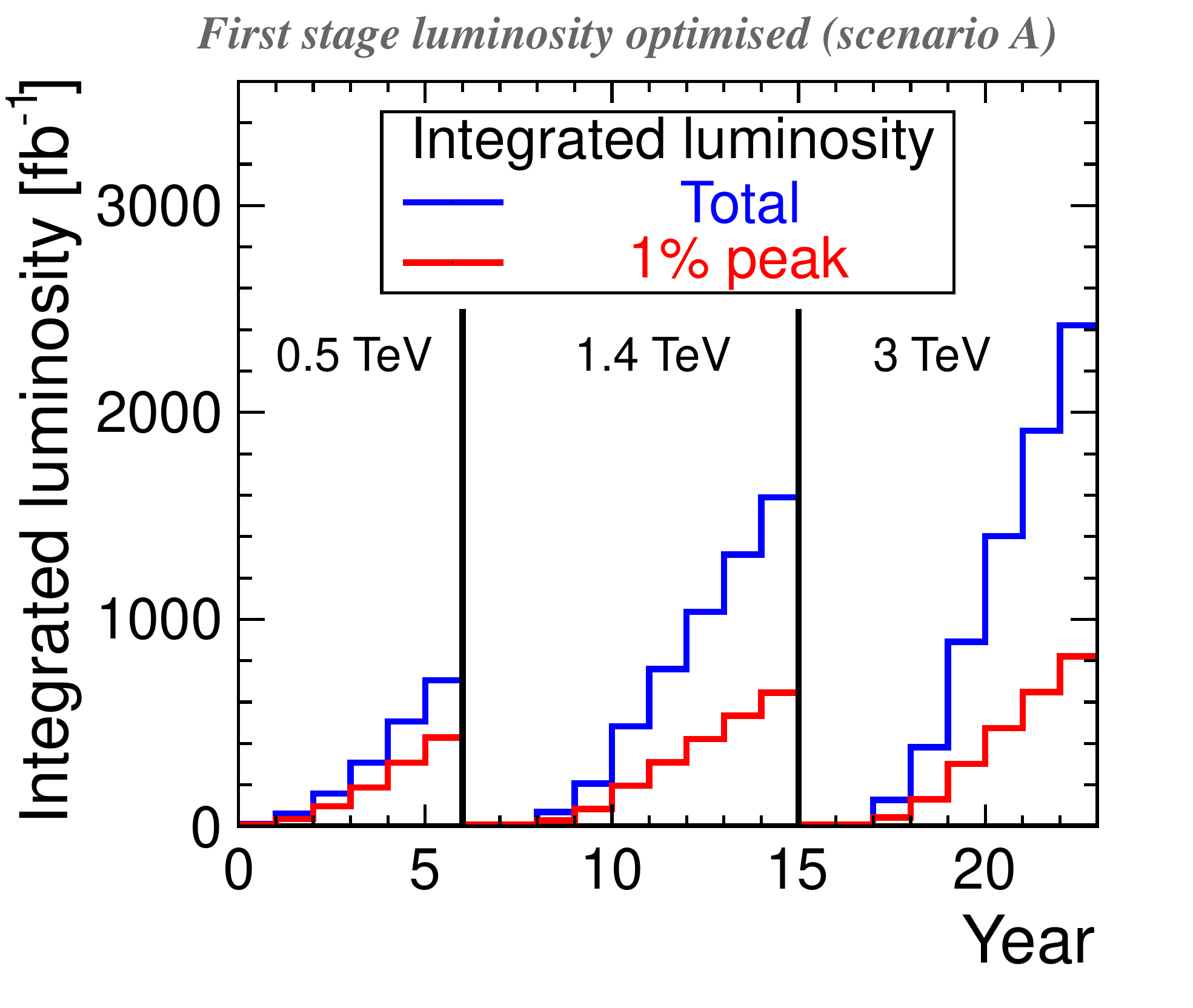}
  \includegraphics[width=0.45\textwidth]{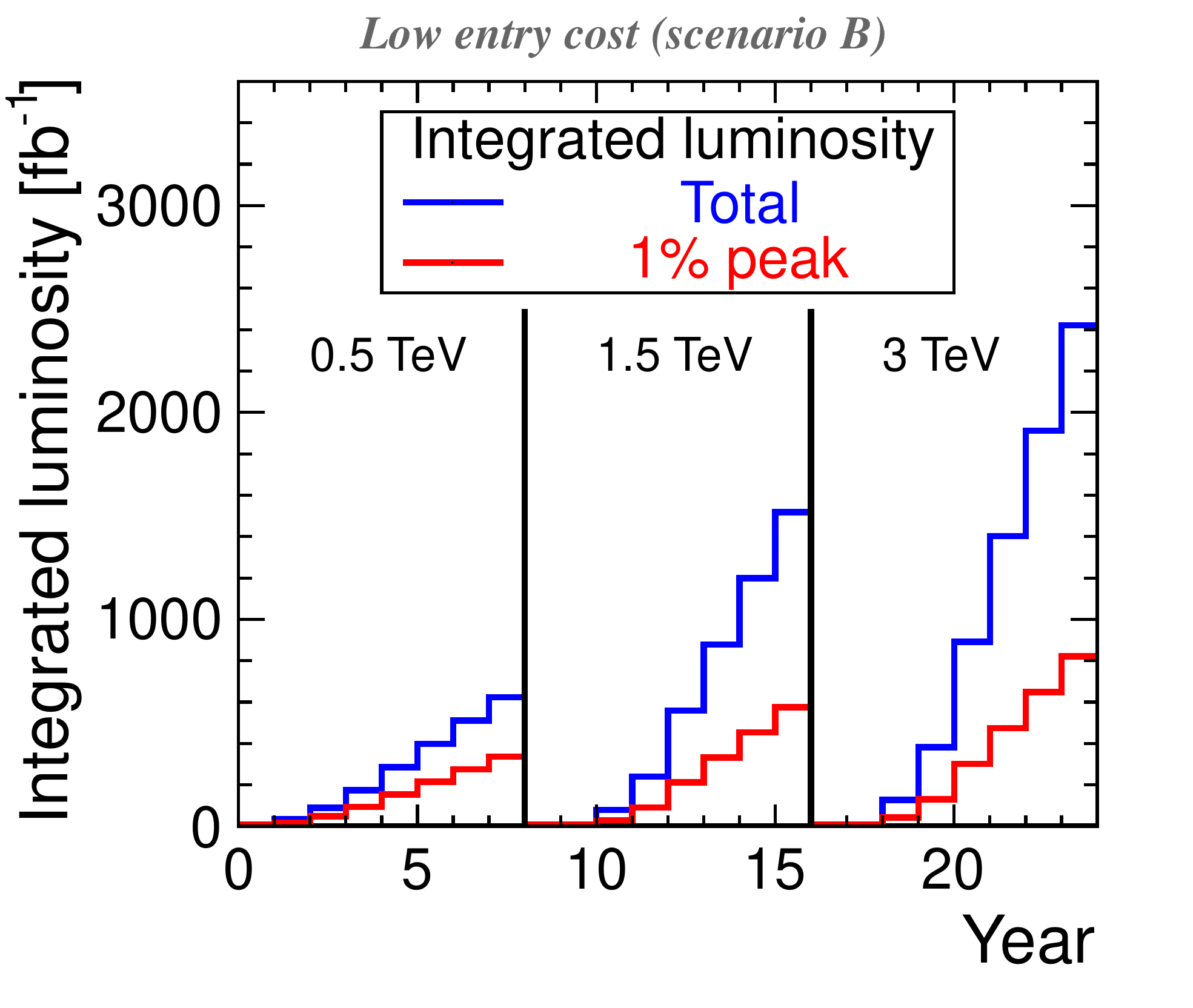}
  \caption{Integrated luminosity in the scenarios optimised for
    luminosity in the first energy stage (left) and optimised for
    entry costs (right). Years are counted from the start of beam
    commissioning.  These figures include luminosity ramp-up of four
    years (5\%, 25\%, 50\%, 75\%) in the first stage and two years
    (25\%, 50\%) in subsequent stages.}
  \label{fig:integratedLumi}
\end{figure}

Possible operating scenarios for the complete CLIC programme
are sketched in Figure~\ref{fig:integratedLumi}: the duration of each
stage is defined by the integrated luminosity targets of 500 fb$^{-1}$
at 500 GeV, 1.5 ab$^{-1}$ at 1.4~(1.5)~TeV and 2~ab$^{-1}$ at~3 TeV
collision energy. The integrated luminosity in the first stage can be
obtained for scenario B by operating for two more years; this is
partly regained in the next stage, so that the overall duration of the
three-stage programme is comparable for both cases, about 24 years
from start of operation.

\begin{figure}[ht!]
  \centering
  \includegraphics[width=\textwidth]{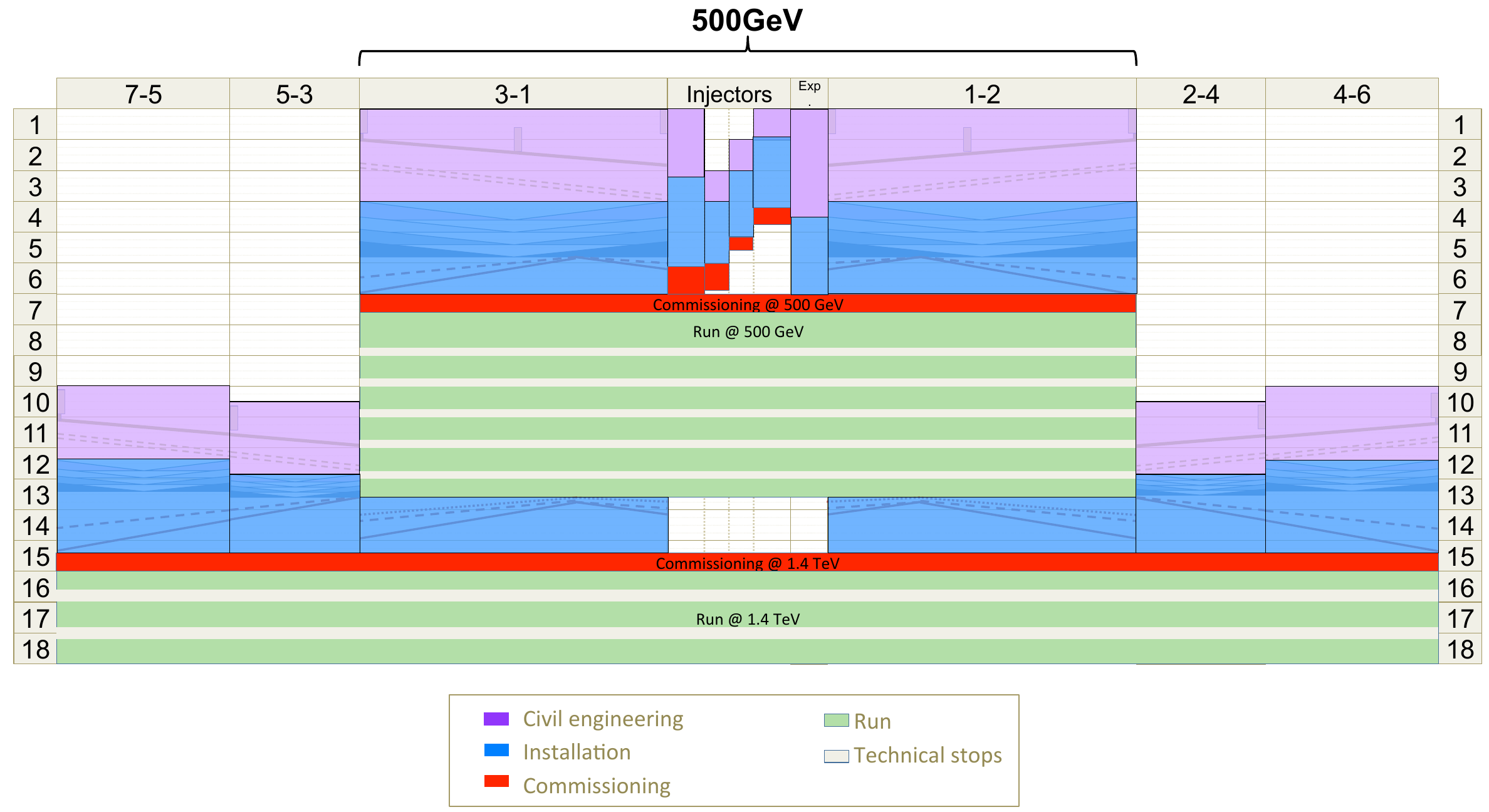}
  \caption[]{Overall ``railway'' schedule for the first two stages of
    scenario A.  The horizontal scale is proportional to tunnel
    length, with the experimental area in the centre. The vertical
    scale shows years from the start of construction.  The
    construction schedule for the main-beam, the drive-beam injectors
    and the experimental area are shown in the centre.}
  \label{fig:overallSchedule_scenarioA}
\end{figure}

Construction schedules (Figure~\ref{fig:overallSchedule_scenarioA})
are essentially driven by civil engineering, infrastructure and
machine installation.
Production of the large-series components proceeds at rates such that
they become available for installation as soon as preceding
construction activities allow it. In the first stage, construction of
the injector complex, experimental area and detectors just matches the
construction time for the main linacs, thus allowing commissioning
with beam to start in year~7.
In order to minimize interruption of operation for physics, civil
engineering and series component production for the second stage must
re-start in year~10, thus allowing commissioning in year~15 (scenario
A): this can be achieved without interference with operation for
physics in the first stage.

The nominal electrical power consumption of all accelerator systems
and services, including the experimental area and the detectors and
taking into account network losses for transformation and distribution
on site, is given in Table~\ref{tab:power} for staging scenarios A and
B.  The table also shows residual power consumption without beams for
two modes corresponding to short ("waiting for beam") and long
("shutdown") beam interruptions. The large variations and volatility
of power consumption will allow CLIC to be operated as a peak-shaving
facility, matching the daily and seasonal fluctuations in power demand
on the network.  Several paths aiming at reducing power consumption or
improving the energy footprint of the machine have been identified and
are under investigation, e.g., reduction of design current density in
magnet windings and cables, replacement of normal-conducting by
permanent or superferric magnets, development of high-efficiency
klystrons and modulators, recovery and valorisation of waste heat.

\begin{figure}[ht!]
  \centering
  \includegraphics[width=0.7\textwidth]{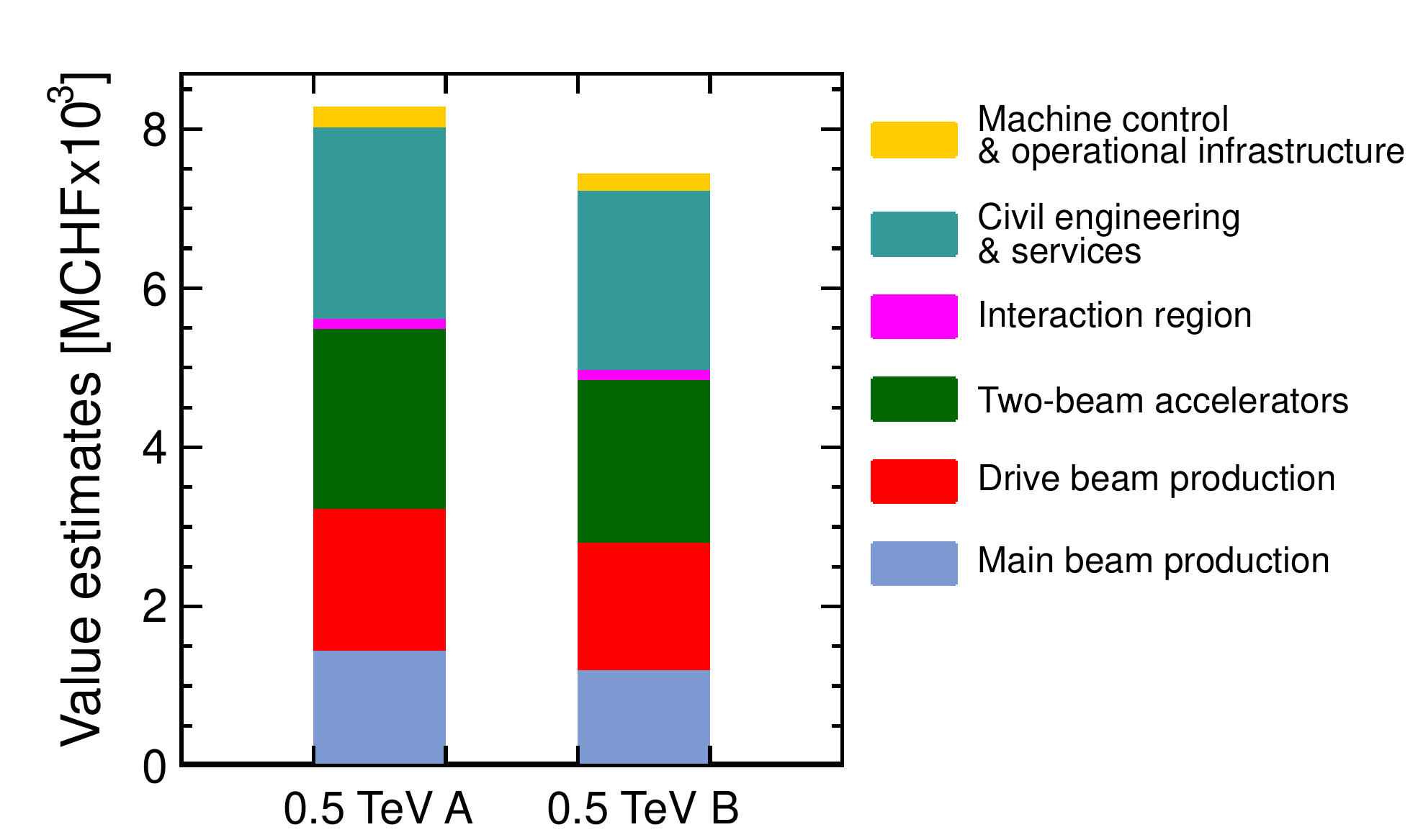}
  \caption{Cost structure of the CLIC accelerator complex at 500 GeV
    for scenarios A and B.}
  \label{fig:cost_500GeV}
\end{figure}

\begin{table}[ht!]
  \caption{\label{tab:power}CLIC power consumption for staging
    scenarios A and B.}
  \centering
  \begin{tabular}{c l l l l }
    \toprule
    Staging scenario &
    $\roots$ [TeV] &
    $P_{\textrm{nominal}}[\textrm{MW}]$ &
    $P_{\textrm{waiting for beam}}[\textrm{MW}]$ &
    $P_{\textrm{shutdown}}[\textrm{MW}]$ \\
    \midrule \midrule
    &  0.5 & 272 & 168 & 37\\
    A & 1.4 & 364 & 190 & 42\\
    &  3.0 & 589 & 268 & 58\\
    \midrule
    &  0.5 & 235 & 167 & 35\\
    B &  1.5 & 364 & 190 & 42\\
    &  3.0 & 589 & 268 & 58\\
    \bottomrule
  \end{tabular}
\end{table}
\begin{table}[ht!]
  \caption{\label{tab:cost}Value and labour estimates of CLIC 500 GeV.}
  \centering
  \begin{tabular}{c l l }
    \toprule
    Staging scenario &
    Value [MCHF] &
    Labour [FTE years]\\
    \midrule \midrule
    A  & $8300^{+1900}_{-1400}$ & 15700 \\[4pt] 
    B  & $7400^{+1700}_{-1300}$ & 14100 \\
    \bottomrule
  \end{tabular}
\end{table}
The cost estimates
follow the ``value'' and ``explicit labour'' methodology used for the ILC
Reference Design report~\cite{ILC_RDR_vol3}.  They are based on the
work breakdown structures established for the different stages of the
two scenarios, and on unit costs obtained for other similar supplies
or scaled from them, and from specific industrial
studies. Uncertainties include technical and procurement risks, the
latter being estimated from a statistical analysis of procurement for
the LHC. The value estimates are expressed in Swiss francs (CHF) of
December 2010 and can thus be escalated using relevant Swiss official
indices. Explicit labour is estimated globally by scaling from LHC
experience. The results are given in Table~\ref{tab:cost}. The cost
structure of the accelerators at 500~GeV collision energy for staging
scenarios A and B is illustrated in Figure~\ref{fig:cost_500GeV}. The
incremental value from the first to the second stage is about 4
MCHF/GeV (scenario B).  Potential savings have been identified for a
number of components and technical systems, amounting to about 10\% of
the total value.  Examples of such savings are the substitution of the
hexapods for the stabilisation of the main-beam quadrupoles with beam
steering, the doubling in length of the support girders for the
two-beam accelerator modules, or the alternative of using assembled
quadrants instead of stacked disks for construction of the
accelerating structures.  Moreover, significant additional savings are
expected from re-optimising the design of the chosen
energy stages.

\section{Detectors}
The detector requirements and resulting layout proposals discussed in the
following are based on the final 3~TeV accelerator stage, which
constitutes the most challenging environment for the detectors.

\textbf{Detector Requirements}\\
The performance requirements for the detector systems at CLIC are
driven by the physics goals described in Section~\ref{sec:physics}.
The
jet-energy resolution should be adequate to distinguish between di-jet
pairs originating from $Z$, $W$ or $H$ bosons. 
This can be achieved with a resolution of 
$\sigma_E/E\sim  3.5\% - 5\%$
for jet energies from 1~TeV down to 50~GeV. The momentum-resolution
requirement for the tracking systems is driven by the precise
measurement of leptonic final states, e.g., the Higgs mass measurement
through $Z$ recoil, where $Z^0\rightarrow \mu^+ \mu^-$, or the determination of slepton
masses in SUSY models. This leads to a required resolution of 
$\sigma_{p_T}/p_T^2 \sim 2 \times 10^{-5} \mathrm{GeV}^{-1}$. 
High-resolution pixel vertex detectors are required
for efficient tagging of heavy quarks through displaced vertices, with
an accuracy of approximately 5~\micron for determining the transverse
impact parameters of high-momentum tracks and a multiple scattering
term of approximately 15~\micron. The latter requires a very low
material budget at the level of $<0.2\%$ of a radiation length per
detection layer, corresponding to a thickness equivalent to less 
than 200~\micron of
silicon, shared by the active material, the readout, the support and
the cooling infrastructure. 

The time structure of the collisions, with bunch crossings spaced by
only 0.5~ns, in combination with the expected high rates of
beam-induced backgrounds, poses challenges for the design of
the detectors and their readout systems. At most one
interesting physics event per 156 ns bunch train is expected, overlaid
by an abundance of particles originating from two-photon
interactions. These background particles will lead to large
occupancies in the inner and forward
detector regions and will require time stamping at the 1--10~ns
level in most detectors, as well as sophisticated pattern-recognition
algorithms to disentangle physics from background events. The gap of
20~ms between consecutive bunch trains will be used for trigger-less
readout of the entire train. Furthermore, most readout subsystems will
be operated in a power-pulsing mode with the most power-consuming
components switched off during the 20~ms gaps, thus taking advantage
of the low duty cycle of the machine to reduce the required cooling
power.

\textbf{Detector Concepts}\\
The detector concepts ILD~\cite{ildloi:2009} and 
SiD~\cite{Aihara:2009ad} developed for the
International Linear Collider (ILC) at a centre-of-mass energy
of 500 GeV form the starting point for the two general-purpose
detector concepts {CLIC\_ILD} and {CLIC\_SiD}. Both detectors will be
operated in one single interaction region in an alternating mode,
moving in and out every few months through a so-called push-pull
system. The main CLIC-specific adaptions to the ILC detector concepts
are an increased hadron-calorimeter depth to improve the containment
of jets at the CLIC centre-of-mass energy of up to 3~TeV and a
redesign of the vertex and forward regions to mitigate the effect of
high rates of beam-induced backgrounds. 

\begin{figure}[ht!] 
        \centering
        \includegraphics[scale=1,clip]{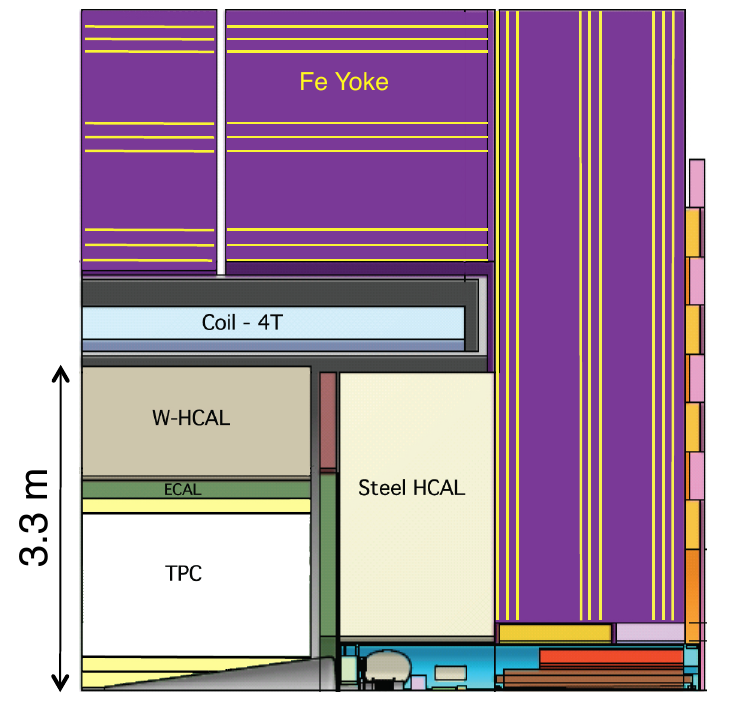}
        \hfill
        \includegraphics[scale=1,clip]{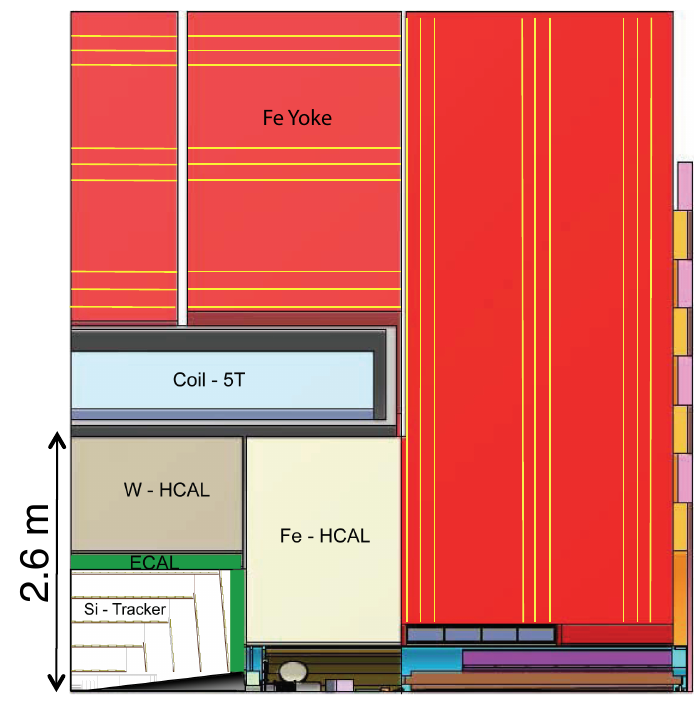}
       \caption{Longitudinal cross-section of the top quadrant of \clicild (left)
        and \clicsid (right). }
        \label{fig:xzDetectorView}
\end{figure}

Figure~\ref{fig:xzDetectorView} shows cross-section views of
{CLIC\_ILD} 
and {CLIC\_SiD}. Both
detectors have a barrel and endcap geometry with the barrel
calorimeters and tracking systems located inside a superconducting
solenoid providing an axial magnetic field of 4~T in case of {CLIC\_ILD}
and of 5~T in case of {CLIC\_SiD}. 

The highly granular electromagnetic and
hadronic calorimeters (ECAL/HCAL) of both detectors are designed for the concept
of particle-flow calorimetry, allowing one to reconstruct individual
particles combining calorimeter and tracking information and thereby
improving the jet-energy resolution to the required unprecedented
levels. The total combined depth of the ECAL and HCAL
is about 8.5 hadronic interaction lengths, realised by changing the
HCAL absorber material from steel to tungsten in the barrel layers.

In the {CLIC\_ILD} concept, the tracking system is based on a large Time
Projection Chamber ({TPC}) with an outer radius of 1.8 m complemented
by an envelope of silicon strip detectors and by a silicon pixel
vertex detector. The all-silicon tracking and vertexing system in
{CLIC\_SiD} is more compact with an outer radius of 1.3 m.

The vertex detectors foresee semiconductor technology with
pixels of approximately $20~\micron~\times~20~\micron$ size.
In case of CLIC\_ILD, both the barrel and
forward vertex detectors consist of three double layers,
while for {CLIC\_SiD}, a geometry with
five single barrel layers and seven single forward layers was chosen. 
The high rates of incoherently produced electron-positron pair-background
events constrain the radius of the central beam pipes
to 29~mm for CLIC\_ILD and to 25~mm for CLIC\_SiD.
For the initial 500~GeV machine, the lower background rates allow for modified vertex-detector
geometries with reduced inner radii.

The superconducting solenoids are surrounded by instrumented iron
yokes allowing one to measure punch-through from high-energy hadron
showers and to identify muons. Two small electromagnetic calorimeters
cover the very forward regions down to 10~mrad. They are foreseen for
electron tagging and for an absolute measurement of the luminosity
through Bhabha scattering.

Full detector simulation studies with event reconstruction demonstrate that both detector
proposals meet the performance requirements and that physics
observables can be measured to high precision.

Preliminary value estimates aiming for an uncertainty of $\pm$30\% and following
the general methodology used for the accelerators place the
CLIC\_ILD detector at 
560~MCHF
and the CLIC\_SiD detector at 
360~MCHF,
excluding explicit labour\footnote{The preliminary CLIC
  detector value estimates were extrapolated, for their major part,
  from the ILC LoI cost estimates, taking the significant changes
  (technology, dimensions) for CLIC into account and using modified
  unit costs.
  Therefore they cannot be directly compared with the ILC
  estimates.}. Main cost drivers for both concepts are the
cost of silicon sensors for the ECAL, and of tungsten for the HCAL.

\begin{figure}[ht!]
\centering
\includegraphics[width=0.49\linewidth]{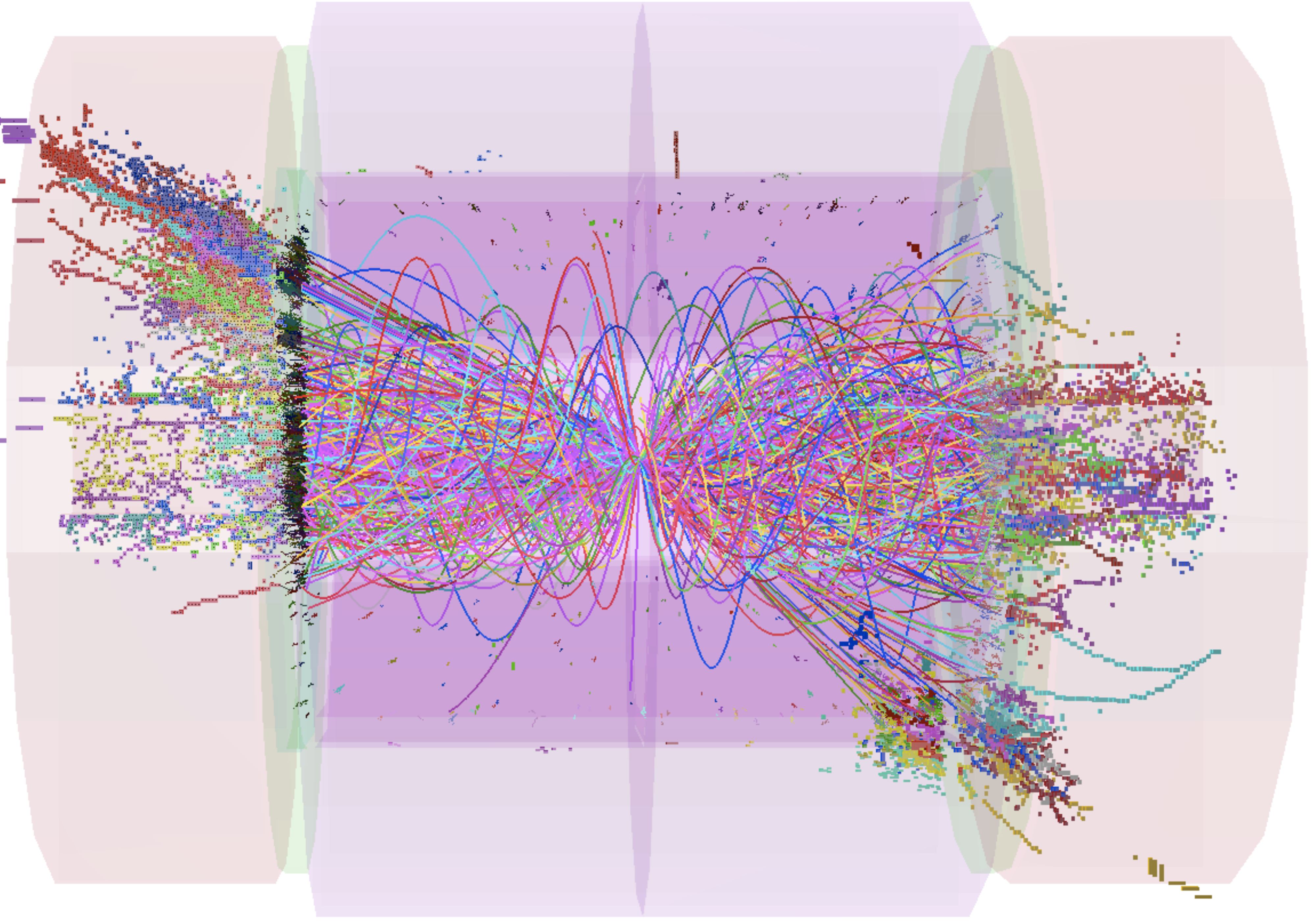}
\includegraphics[width=0.49\linewidth]{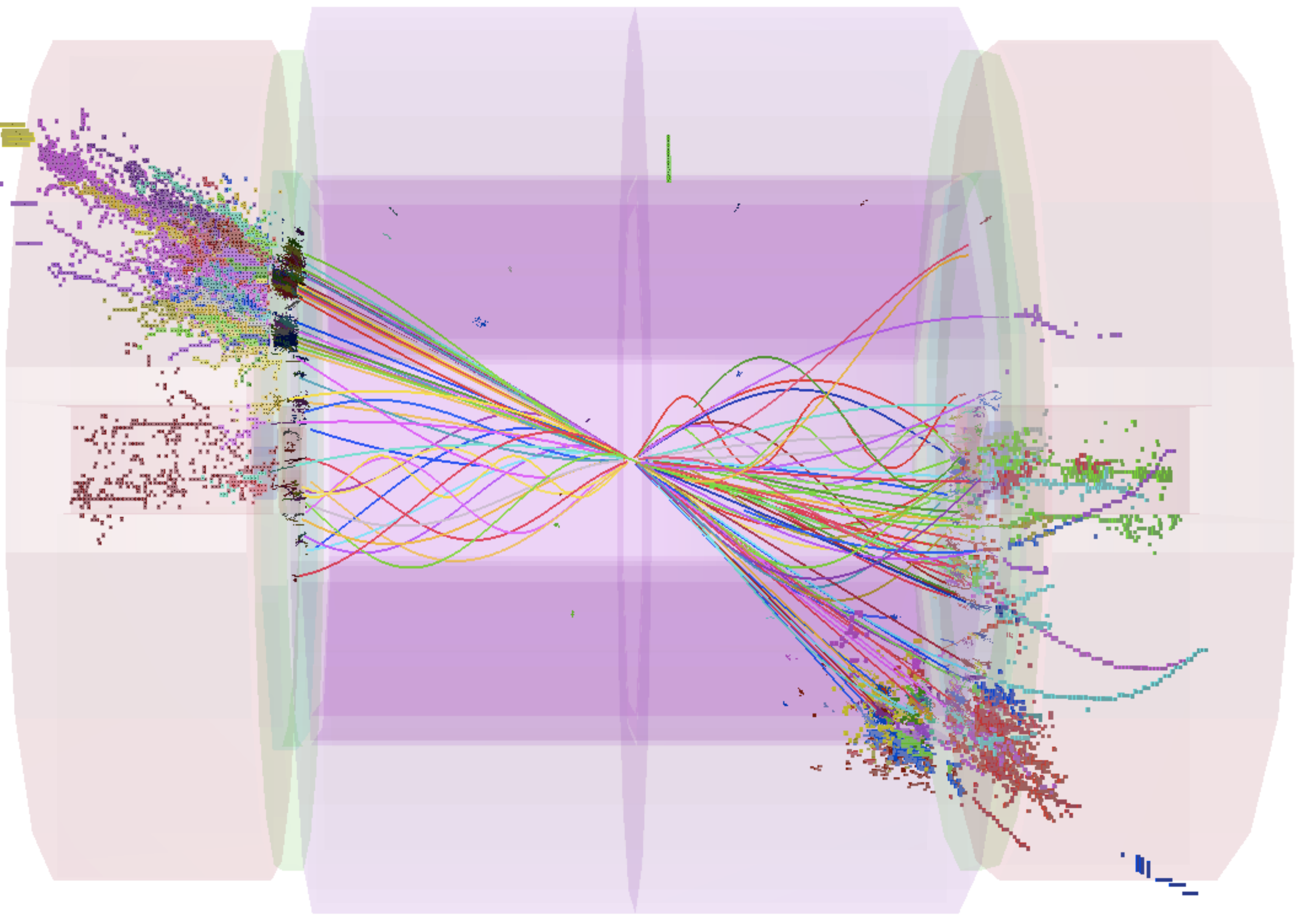}
\caption{Left: Reconstructed particles in a simulated
  $e^+e^-\rightarrow t\bar{t}$ event at 3~TeV in the \clicild detector
  concept with background from $\mathrm{\gamma\gamma \to\textrm{hadrons}}$ overlaid. Right: The effect
  of applying tight timing cuts on the reconstructed cluster times.}
\label{fig:ttEventDisplay}
\end{figure}

\textbf{Suppression of Beam-induced Background}\\
Even at 3~TeV, the high levels of beam-induced background can be
suppressed by making use of the high spatial and temporal granularity
provided by the detectors. For this, a scheme
was developed, which considers time stamping capabilities of 10~ns
for all silicon tracking elements, and of 1~ns time resolution for all
calorimeter hits.
Broad timing cuts around the time of the physics event, identified offline,
are followed by a tighter set of cuts applied to reconstructed
low-$p_T$ particle-flow objects.
As a result, the average background level can be reduced from
approximately 20~TeV per bunch train to
about 100~GeV per reconstructed physics event. This 
background rejection, which is exemplified in
Figure~\ref{fig:ttEventDisplay}, is achieved without significantly
impacting the detector performance. 

\section{Physics Potential}
\label{sec:physics}

\begin{figure}[ht!]
\begin{minipage}[c]{0.4\linewidth}
\centering
\includegraphics[scale=0.39]{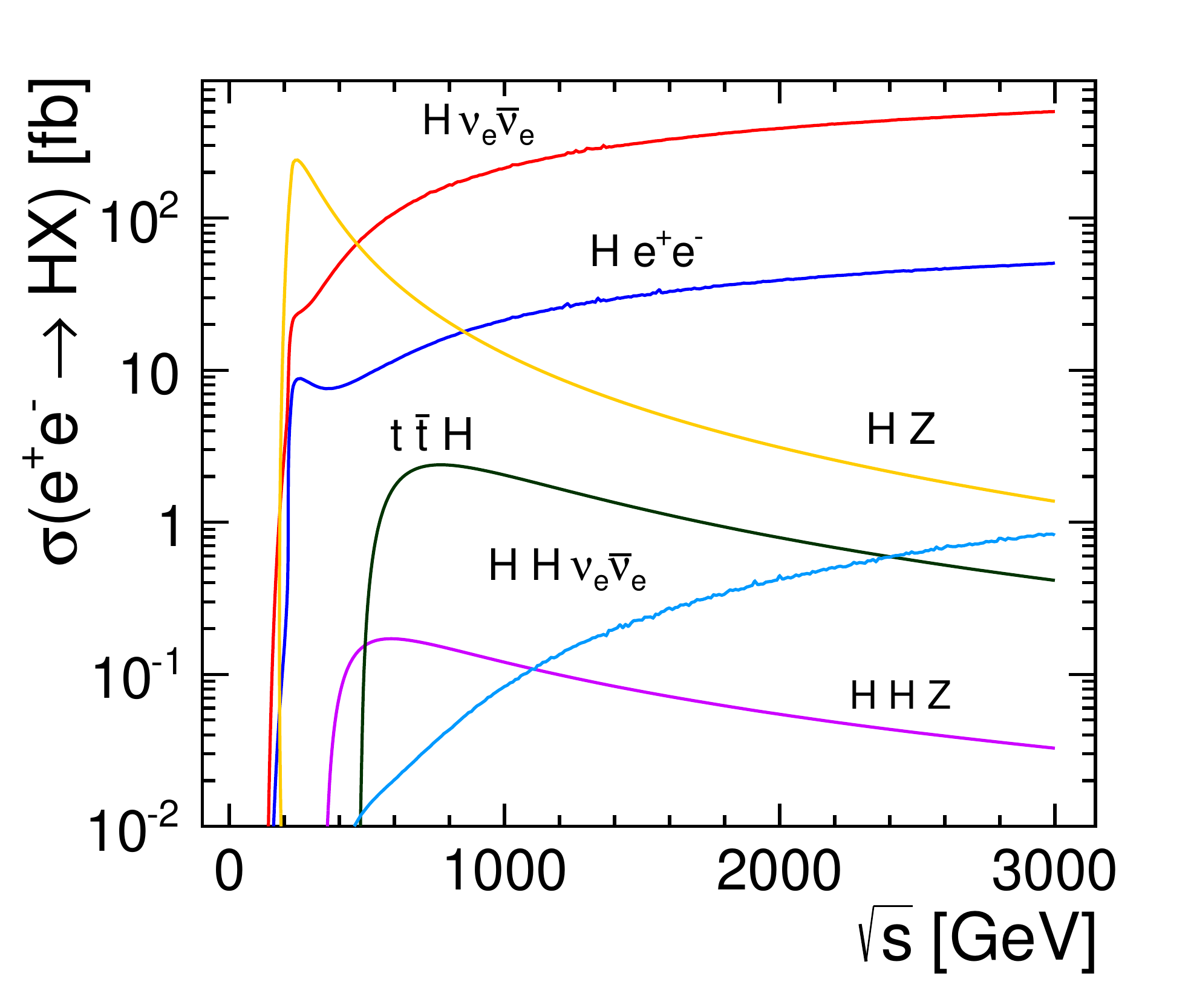}
\end{minipage}
\hspace{0.5cm}
\begin{minipage}[c]{0.6\linewidth}
\centering
\includegraphics[scale=0.41]{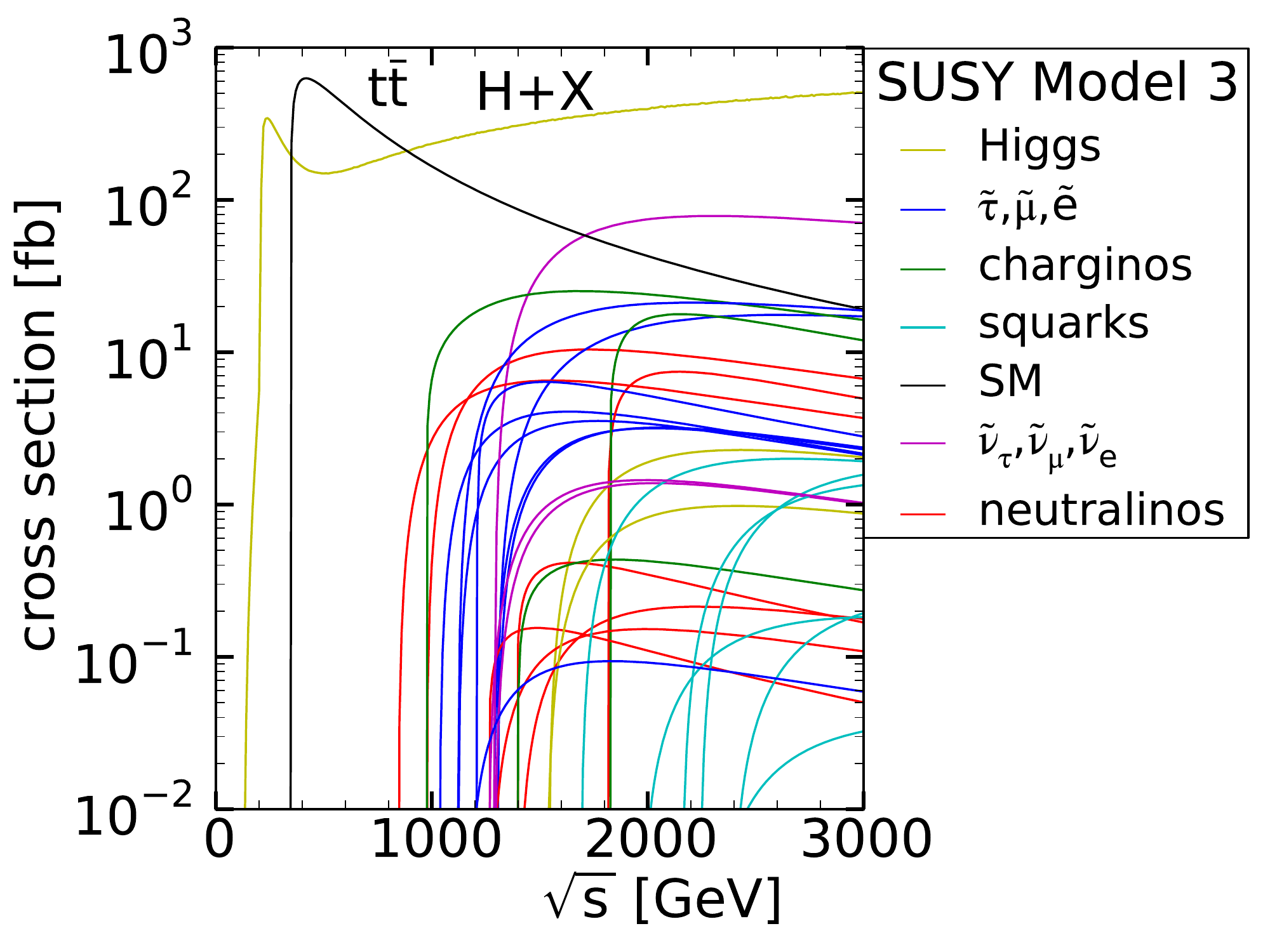}
\end{minipage}
\caption{Left: Production cross-sections of the SM Higgs boson in
  \epem collisions as a function of $\roots$ for $m_H=125$~GeV. 
  Right: SUSY production cross-sections of \modelthree as a function of $\roots$. 
  Every line of a given colour corresponds to the production cross
  section of one particle in the legend.}
\label{fig:higgs_modelIII_xsec}
\vspace{5mm}
\end{figure}

Recently the ATLAS and CMS collaborations have announced evidence for
a state consistent with a SM Higgs boson with a mass of 125~GeV. The
production cross-sections as a function of \epem centre-of-mass energy
of a SM Higgs boson of that mass is given in
Figure~\ref{fig:higgs_modelIII_xsec} (left). The cross-sections are in
the hundreds of fb. Therefore, tens of thousands of events  can be
obtained with hundreds of ${\rm fb}^{-1}$ of integrated luminosity,
which is anticipated for CLIC running.  

CLIC is able to measure this boson's couplings to SM states with extraordinary precision. A summary of Higgs observables and the precision with which they can be determined is provided in Table~\ref{tab:Higgs}. These numbers are obtained from studies that employ full detector simulations with backgrounds overlaid. 
For several standard channels the statistical error is at the percent level.
For other channels, such as the low-rate $H\to \mu^+\mu^-$ and the Higgs pair production,  both of which will be severe challenges for the LHC, the statistical error is near the 20\% level.  These levels of precision, which generally go well beyond LHC capabilities, are needed to resolve the subtle shifts in Higgs boson couplings that are present in many beyond the SM theories.

\begin{table}[ht!]
\centering

\caption{Summary of results obtained in the Higgs studies for $m_{H}=$120~GeV. All analyses at centre-of-mass energies of 350~GeV and 500~GeV assume an integrated luminosity of 500~$\textrm{fb}^{-1}$, while the analyses at 1.4 TeV (3~TeV) assume 1.5~$\textrm{ab}^{-1}$(2~$\textrm{ab}^{-1}$).}
\label{tab:Higgs}
\begin{tabular}{l l l l l l l l}
  \toprule
  \multicolumn{8}{c}{\textbf{Higgs studies for $m_{H}=$120~GeV}}\\\midrule
  $\roots$             & \multirow{2}{*}{Process} & Decay                                                 & Measured        & \multirow{2}{*}{Unit} & Generator & Stat. & \multirow{2}{*}{Comment}\\
  (GeV)                &                          & mode                                                  &                 quantity            &                       & value     & error   \\\midrule\midrule

  \multirow{3}{*}{350} &                          & \multirow{3}{*}{$ZH\rightarrow \mu^+\mu^-X$}          & $\sigma$                           & fb                    & 4.9       & 4.9\% & Model \\ \cmidrule{4-7}
                       &                          &                                                       & Mass                               & GeV                   & 120       & 0.131 & independent,\\
                       &                          &                                                       &                                    &                       &           &       & using $Z$-recoil\\\cmidrule{1-1}\cmidrule{3-8}

  \multirow{3}{*}{500} &  SM Higgs                & \multirow{3}{*}{$ZH\rightarrow q\bar{q}q\bar{q}$}     & $\sigma \times$ BR                 & fb                    & 34.4       & 1.6\% & $ZH\rightarrow q\bar{q}q\bar{q}$  \\\cmidrule{4-7}
                       & production               &                                                       & Mass                               & GeV                   & 120       & 0.100 & mass  \\
                       &                          &                                                       &                                    &                       &           &       & reconstruction\\ \cmidrule{1-1}\cmidrule{3-8}

  \multirow{2}{*}{500} &                          &$ZH,H\nu\bar{\nu} $ & $\sigma \times$ BR                 & fb                    & 80.7      & 1.0\% & Inclusive  \\\cmidrule{4-7}
                       &                          &                                   $\rightarrow \nu\bar{\nu}q\bar{q}$                    & Mass                               & GeV                   & 120       & 0.100 & sample \\ \midrule

  1400                 &                          & $H\rightarrow \tau^+\tau^-$                           &\multirow{4}{*}{$\sigma \times$ BR} & \multirow{4}{*}{fb}   & 19.8      & $<$3.7\% & \\\cmidrule{1-1}\cmidrule{3-3}\cmidrule{6-7}
 \multirow{3}{*}{3000} &  $WW$                    & $H\rightarrow b\bar{b}$                               &                                    &                       & 285      & 0.22\% & \\
                       &  fusion                  & $H\rightarrow c\bar{c}$                               &                                    &                       & 13        & 3.2\% & \\
                       &                          & $H\rightarrow \mu^+\mu^-$                             &                                    &                       & 0.12      & 15.7\% & \\\midrule

                       &                          &                                                       & Higgs                              &  & &  & \\
  1400                 & $WW$                     &                                                       & tri-linear                         &                       &   &  $\sim$20\% & \\ 
  3000                 & fusion                   &                                                       & coupling                           &                       &   &       $\sim$20\%        & \\  
                       &                          &                                                       & $g_{HHH}$                          &                       &   &              & \\  

 \bottomrule \vspace{2mm}
\end{tabular}
\end{table}


Being the heaviest particle in the SM, the top quark couples more
strongly to the Higgs boson than any other fermion in the SM. The
precise knowledge of the properties of the top quark also provides
important sensitivity to physics beyond the SM.
Normal kinematic determinations
of the top quark mass are beset by significant QCD uncertainties when
trying to map it to 
a theoretically well-defined mass definition.
These subtleties
can be overcome most easily by performing a measurement of the $t\bar
t$ cross-section at multiple points near
threshold. Table~\ref{tab:top} shows the results of two threshold
scans, one of 6 points and one of 10 points, each separated by 1~GeV
with $10~ {\rm fb}^{-1}$ of luminosity at each point.
In the latter scan the value of $\alpha_s$ is allowed to vary in the fit
and is determined simultaneously with the top quark mass from the scan
data.  A third row shows the invariant mass measurement from $100~
{\rm fb}^{-1}$ of integrated luminosity at $\sqrt{s}=500$~GeV. 
\begin{table}[tb]
\centering
\caption{Summary of full detector-simulation results obtained under
 realistic CLIC beam conditions in the top quark studies. 
The first (second) threshold scan contains 6 points (10 points) separated
by 1~GeV and with $10\, {\rm fb}^{-1}$ of luminosity at each point.}
\label{tab:top}
\begin{tabular}{l l l l l l l}
  \toprule
  \multicolumn{7}{c}{\textbf{Top studies}}\\\midrule
  $\roots$ & \multirow{2}{*}{Technique} &Measured & Integrated             & \multirow{2}{*}{Unit} & Generator & Stat. \\
  (GeV)    &                              &             quantity       &  luminosity ($\fbinv$) &                       & value     & error   \\\midrule\midrule
  
  \multirow{3}{*}{350} &  \multirow{3}{*}{Threshold scan} & Mass & $6\times 10$ & GeV & 174 & 0.021 \\ \cmidrule{3-7}
                       &                                  & Mass & \multirow{2}{*}{$10\times 10$} & GeV & 174 & 0.033 \\
                       &                                  & $\alpha_S$ &                          &     & 0.118 & 0.0009\\\midrule

  500                  &  Invariant mass                  & Mass & 100 & GeV & 174 & 0.060 \\

  \bottomrule
\end{tabular}
\end{table}

The Higgs boson and top quark measurements form a core component of
the physics programme for a staged CLIC collider. Going into the energy
frontier there are opportunities to discover new physics and to do
precision studies of potential discoveries at the LHC. A useful
example to study in this context is supersymmetry, since it is a
well-motivated idea and the particle content is rich, calculable and
illustrative of many new physics ideas that have new particles. 
In Figure~\ref{fig:higgs_modelIII_xsec} (right) we show the production
cross-section spectrum of an example supersymmetry scenario
(\modelthree~\cite{CLICCDR_vol3})
as a function of the centre-of-mass
energy. One sees the light Higgs boson and top quark
at lower energies. At higher energies the electroweak gaugino
and slepton thresholds open up, which could be accessible at a second
intermediate stage. At still higher energies some of the strongly
interacting squark thresholds become accessible, as well as the heavy
Higgs bosons. When kinematically accessible they can be measured with
excellent precision.

As an example of how well new particles can be discovered and their
masses and interactions precisely studied, we present in
Table~\ref{tab:susy-maccuracy} the results of simulation studies of
supersymmetry \modeltwo~\cite{CLICCDR_vol2}. The table shows the
masses of these various sparticle states, associated to this model, and
the statistical accuracy by which they could be determined at 3~TeV
CLIC with $2~ {\rm ab}^{-1}$ of integrated luminosity. In addition,
Table~\ref{tab:susy_summaryTable_14} shows the results of full
simulation benchmark studies including background overlay for various processes relevant to
supersymmetry \modelthree at 1.4~TeV CLIC. In both cases,  
many of the determinations are at the percent level of accuracy or
better. The precision of these measurements are by no means excessive
-- they are crucial for making distinctions between various underlying
supersymmetry breaking and transmission scenarios that have different
unification characteristics at the high scale (e.g., the grand unified
scale).

\begin{table}[ht!]
\caption{Values of the SUSY particle masses of the chosen benchmark point (\modeltwo)
and estimated experimental statistical accuracies at CLIC, as obtained
in the analyses presented in Chapter~12 of~\cite{CLICCDR_vol2}, and also
in~\cite{Accomando:2004sz} (indicated with $^*$). All values are in~GeV. 
The last column is either out of kinematic reach or not studied.  
All studies are performed at a centre-of-mass energy 
of \unit[3]{TeV} and for an integrated luminosity of \unit[2]{\abinv}.
}
\begin{center}
\begin{tabular}{crr}
\toprule
Particle               & Mass   & Stat.\ acc.  \\
\midrule
$\widetilde{\chi}_1^0$     & 340.3  & $\pm3.3\;\;$ \\
$\widetilde{\chi}_2^0$     & 643.1  & $\pm9.9\;\;$ \\

$\widetilde{\chi}_3^0$     & 905.5  & $\pm$19.0$^*$\\

$\widetilde{\chi}_4^0$     & 916.7  & $\pm$20.0$^*$\\
$\widetilde{\chi}_1^{\pm}$ & 643.2  & $\pm3.7\;\;$ \\

$\widetilde{\chi}_2^{\pm}$ & 916.7  &$\pm$7.0$^*$  \\
$\widetilde{e}_R^{\pm}$    & 1010.8 & $\pm2.8\;\;$ \\
$\widetilde{\mu}_R^{\pm}$  & 1010.8 & $\pm5.6\;\;$ \\
$\widetilde{\nu}_1$        & 1097.2 & $\pm3.9\;\;$ \\
\bottomrule
\end{tabular}
\quad
\begin{tabular}{crr}
\toprule
Particle  & Mass  & Stat.\ acc.   \\
\midrule

$h$       & 118.5 & $\pm $0.1$^*$ \\
$A^0$       & 742.0 & $\pm 1.7\;\;$ \\
$H^0$       & 742.0 & $\pm 1.7\;\;$ \\
$H^{\pm}$ & 747.6 & $\pm2.1\;\;$  \\
\bottomrule
\toprule
Quantity          & Value & Stat.\ acc.\\
\midrule
$\Gamma(A^0)$       & 22.2  & $\pm3.8$~ \\
$\Gamma(H^{\pm})$ & 21.4  & $\pm4.9$~ \\
\bottomrule
\end{tabular}
\quad
\begin{tabular}{cr}
\toprule
Particle         & Mass   \\
\midrule
$\widetilde{\tau}_1$ & $670$  \\
$\widetilde{\tau}_2$ & $974$  \\
$\widetilde{t}_1$    & $1393$ \\
$\widetilde{t}_2$    & $1598$ \\
$\widetilde{b}_1$    & $1544$ \\
$\widetilde{b}_2$    & $1610$ \\
$\widetilde{u}_R$    & $1818$ \\
$\widetilde{u}_L$    & $1870$ \\
$\widetilde{g} $     & $1812$ \\
\bottomrule
\end{tabular}
\end{center}
\label{tab:susy-maccuracy}
\end{table}
\begin{table}[ht!]
\vspace{1cm}
    \centering
    \caption{Summary table of the CLIC SUSY benchmark analyses results
     obtained with full detector simulations with background overlaid.
    All studies are performed at a centre-of-mass energy 
     of \unit[1.4]{TeV} and for an integrated luminosity of \unit[1.5]{\abinv}.\label{tab:susy_summaryTable_14}}
    \begin{tabular}{l l l l l l l l}
        \toprule
        $\roots$ & Process & Decay mode & SUSY & Measured & Unit & Gene- & Stat. \\
        (TeV)    &         &            & model &      quantity     &      &
    rator & uncert- \\
                 &         &            &       &           &      &
    value &  ainty\\
        \midrule
	
	\midrule
        \multirow{9}{*}{1.4} &            &  \multirow{3}{*}{$\widetilde{\mu}_R^+ \widetilde{\mu}_R^-\rightarrow \mu^+\mu^-\widetilde{\chi}_1^0 \widetilde{\chi}_1^0 $} & \multirow{9}{*}{III}  & $\sigma$               & fb  & 1.11  & 2.7\%  \\
	                     &            &                                                                        &                       &  $\tilde\ell$ mass     & GeV & 560.8 &  0.1\% \\
                             &            &                                                                        &                       &  $\widetilde{\chi}_1^0$ mass & GeV & 357.8 & 0.1\%  \\ \cmidrule{5-8}

	                     & Sleptons   & \multirow{3}{*}{$\widetilde{e}_R^+\widetilde{e}_R^-\rightarrow e^+e^-\widetilde{\chi}_1^0 \widetilde{\chi}_1^0$}  &                       & $\sigma$               & fb  & 5.7   & 1.1\%  \\
	                     & production &                                                                        &                       & $\tilde\ell$ mass      & GeV & 558.1 & 0.1\%  \\
                             &            &                                                                        &                       & $\widetilde{\chi}_1^0$ mass  & GeV & 357.1 & 0.1\%  \\\cmidrule{5-8}

	                    &             & \multirow{3}{*}{$\widetilde{\nu}_e \widetilde{\nu}_e\rightarrow \widetilde{\chi}_1^0 \widetilde{\chi}_1^0 e^+e^- W^+W^-$}&                       & $\sigma$               & fb  & 5.6   & 3.6\% \\
	                    &             &                                                                        &                       & $\tilde\ell$ mass      & GeV & 644.3 & 2.5\% \\
                            &             &                                                                        &                       & $\widetilde{\chi}_1^{\pm}$ mass  & GeV & 487.6 & 2.7\% \\
 
	\midrule
	\multirow{2}{*}{1.4} & Stau       & \multirow{2}{*}{$\widetilde{\tau}_1^+\widetilde{\tau}_1^-\rightarrow \tau^+\tau^-\widetilde{\chi}_1^0\widetilde{\chi}_1^0$} & \multirow{2}{*}{III} & $\widetilde{\tau}_1$ mass & GeV & 517 & 2.0\%\\
                             & production &                                                                                                             &                      &$\sigma$               & fb  & 2.4 & 7.5\%\\

	\midrule
	\multirow{4}{*}{1.4} & Chargino   &  \multirow{2}{*}{$\widetilde{\chi}_1^+\widetilde{\chi}_1^-\rightarrow \widetilde{\chi}_1^0 \widetilde{\chi}_1^0  W^+W^-$}        & \multirow{4}{*}{III} & $\widetilde{\chi}_1^{\pm}$ mass& GeV & 487  & 0.2\% \\
	                     & production &                                                                     &                      & $\sigma$             & fb  & 15.3 & 1.3\% \\\cmidrule{2-3}\cmidrule{5-8}
	                     & Neutralino & \multirow{2}{*}{$\widetilde{\chi}_2^0\widetilde{\chi}_2^0\rightarrow h/Z^0 h/Z^0 \widetilde{\chi}_1^0 \widetilde{\chi}_1^0$}&                      & $\widetilde{\chi}_2^{0}$ mass& GeV & 487  & 0.1\% \\
	                     & production &                                                                     &                      & $\sigma$             & fb  & 5.4  & 1.2\% \\

	
        \bottomrule
	\\
	\\ 
    \end{tabular}
\end{table}


\newpage
 A complementary and less model-dependent approach of
   probing physics beyond the SM aims for detecting evidence for
   higher-dimensional operators, or contact interactions, of SM states,
   which may arise from integrating out exotic particles.
Dimension-six operators ${\cal O}^{(6)}/\Lambda^2$ can be probed up to
scales of $\Lambda=60$~TeV, with some variability for each particular
operator and beam polarisation~\cite{CLICCDR_vol2}. The analogous
reach at the LHC with 100~fb${}^{-1}$ is currently projected to be less than 8~TeV.


Finally, we summarise the {CLIC} reach at 3~TeV compared to other
collider options for several New Physics models. The
sensitivity scale for squarks is better at the LHC than at a 3~TeV CLIC,
except for some difficult cases with small mass splittings. 
On the other hand, weakly interacting particles,
such as sleptons, have much higher direct reach at a 3~TeV CLIC than
LHC. The $Z'$ searches are generally up to about 20~TeV at 3~TeV CLIC,
although the details depend on the precise model. The analogous
sensitivity at the LHC is about 5~TeV. Further concepts of new
physics are
catalogued in Table~\ref{fig:chap2:discovreach}, including triple gauge
coupling (TGC) deviations from the SM values, the contact
interaction scale involving electrons and muons (``$\mu$ contact
scale''), and the scale of Higgs compositeness that would result in
detectable shifts in the Higgs boson observables away from their SM
values.

\begin{table}[ht!]
\begin{center}
  \caption{Discovery reach of various theory models for different
colliders and various levels of integrated luminosity
$\mathcal{L}$~\cite{DeRoeck:2001nz}. LHC14 and the luminosity-upgraded
 SLHC are both at \roots=14~TeV and with performance assumptions
which will likely be updated in the context of
this strategy process.
LC800 is an 800~GeV $e^+e^-$ collider and
CLIC3 is at $\roots=3$~TeV. 
TGC is short for Triple Gauge Coupling, and
``$\mu$ contact scale'' is short for LL $\mu$ contact interaction scale 
$\Lambda$ with $g=1$ (see Chapter~1 in~\cite{CLICCDR_vol2} and
references therein).}
\label{fig:chap2:discovreach}
 \begin{tabular}{ llllll }
    \toprule
\multirow{2}{*}{Particle / parameter} & Collider: &    LHC14 & SLHC & LC800 & CLIC3\\
                      & $\mathcal{L}$: &    100~\fbinv & 1~\abinv & 500~\fbinv &
                      1~\abinv \\
\midrule
Squarks [TeV] & &   2.5 & 3 & 0.4 & 1.5 \\
Sleptons [TeV] &  & 0.3 & - & 0.4 & 1.5 \\ 
$Z'$ ({\tiny SM ~couplings}) [TeV]  & &  5 & 7 & 8 & 20   \\ 
2 extra dims $M_D$ [TeV]  &  &   9 & 12 & 5-8.5 & 20-30 \\
TGC (95\%)  ({\tiny \rm $\lambda_{\gamma} $~coupling}) & &   0.001& 0.0006& 0.0004& 0.0001 \\
$\mu$ contact scale [TeV] &  & 15& - & 20 & 60 \\
Higgs compos. scale [TeV] & & 5-7 & 9-12 & 45 & 60\\
    \bottomrule
  \end{tabular}
\end{center}
\end{table}

\vspace{-5mm}
\section{Summary and Outlook}

Linear \epem colliders have an impressive physics potential that is in
many ways complementary to LHC. A comprehensive overview document of
the Linear Collider physics capabilities has been composed by a small
team of nominated experts from the {ILC} and {CLIC} communities and
submitted to the update process of the European Strategy for Particle
Physics~\cite{LC_Phys_Case_Strategy_Input}.  This document outlines
the strong physics case for a {LC}. It has been reviewed and is
supported by the full international {LC} community.

The {CLIC} studies that are relevant for the European Strategy
Process have been carried out in the framework of preparing the {CLIC}
Conceptual Design Report, with two main volumes covering the
accelerator studies~\cite{CLICCDR_vol1} and the physics and detector
studies~\cite{CLICCDR_vol2}, respectively.  These two volumes
demonstrate the feasibility of the {CLIC} accelerator concept in
detail, and show the large physics potential of such an accelerator based on
detailed detector and physics simulations in a realistic experimental
environment. They are particularly focused on the technical challenges
of construction and operation of a 3 TeV {CLIC}-technology-based
accelerator and corresponding detectors. Intermediate energies, in particular an
initial stage at 500 GeV, are also introduced in these volumes.
Several of these
studies have also been carried out in synergy and close collaboration
with the {ILC} studies. 

\begin{figure}[t!]
  \centering
  \includegraphics*[width=\textwidth]{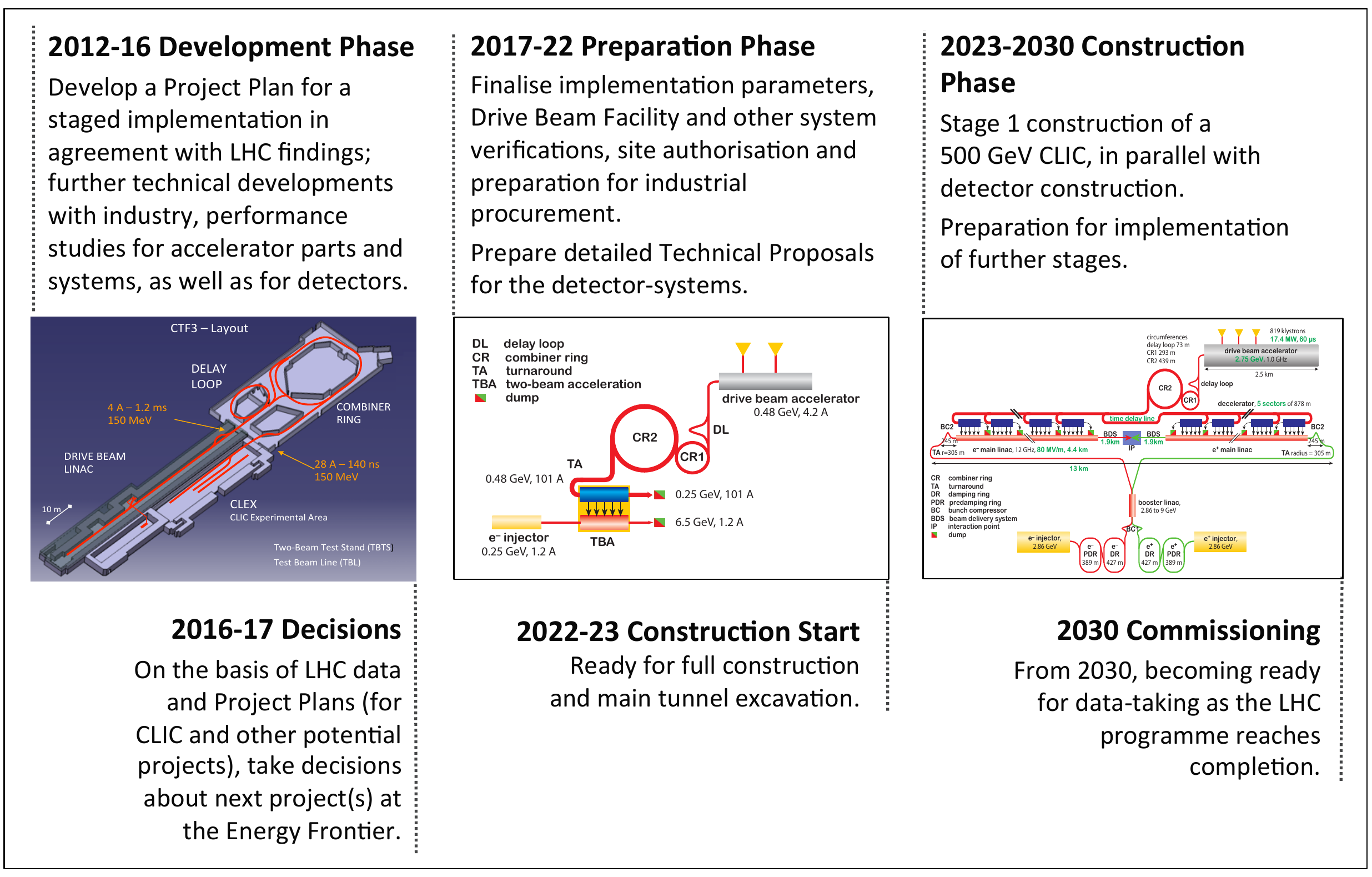}
  \caption{Top row: An outline of the CLIC project timeline with main
    activities leading up to and including the first stage
    construction. Middle row: illustrations of the CTF3 facility (one
    of several testing facilities of importance to the project
    development), a new large drive beam facility with final CLIC 
    elements
    which is also needed for acceptance tests, and a 500 GeV
    implementation.  Bottom row: Main decision points and activity
    changes.} 
  \label{f:timeline}
\end{figure}

Recent work carried out by the {CLIC}
collaboration and the {CLIC} physics and detector study
has also addressed project-implementation issues such as:
site studies, cost and power,
the construction and operation of {CLIC} in three energy stages and
its positive impact on the physics 
potential. These subjects
are described in a third CDR volume~\cite{CLICCDR_vol3} that also
includes summaries of the two 
other CDR reports and forms the basis for this input
to the European Strategy for Particle Physics.

The {CLIC} project as outlined is an ambitious long-term
programme, with an initial 7 year construction period and three energy
stages each lasting 6--8 years, interrupted by 2 year upgrade periods. A
development programme for the {CLIC} project has been established and
is being carried out concurrently with {LHC} operation at 8~TeV and
later full energy, covering the period until 2016. By that time both
the {LHC} physics results and technical developments should have
reached a maturity that would allow a decision about the most
appropriate next project(s) at the energy frontier. The major
contenders, with particular relevance for the European Strategy, are a
Linear Collider or an energy-upgraded {LHC}.

These options can provide a long-term strategy for European Particle
Physics well beyond 2030. They represent investments, commitment and
physics scope well beyond the {LHC} programme, including its
luminosity upgrade, that is likely to remain the main experimental
facility at the energy frontier until ~2030. Construction start for
{CLIC} could be around 2023 after an initial Project Preparation Phase
2017--2022, in time for completion by 2030 when the {LHC} programme
reaches a natural completion.  The currently foreseen timeline for the
CLIC project is shown in Figure~\ref{f:timeline}, with details
presented in ~\cite{CLICCDR_vol3}.
\newpage
With the recent discovery of a new Higgs-like state at
$\sim 125$~GeV at {LHC} and considering the importance of studies
near the top threshold, it is evident that
an initial {CLIC} stage at 400--500~GeV will already provide
exceptional physics. 
A second stage around ~1.2--1.5~TeV would allow
measurements of several more difficult Higgs branching ratios and in
particular the Higgs self-coupling. With the present knowledge
a third stage well beyond
1.5~TeV can only be justified by the general arguments of improved
production cross-sections and precision on the measurements mentioned
above, and a significantly increased search capability.  It is however
important to keep in mind that the very recent results from {LHC} open
a completely new experimental territory. We can look forward to more
{LHC} results during 2012--2013 and when {LHC} moves to full energy
running in 2015, potentially providing even more exciting prospects
for a future {CLIC} programme, including ultimate energy stages
beyond ~1.5 TeV.

\vspace{2mm}\textbf{Messages for the Strategy Process}\\
The feasibility studies for the CLIC accelerator have over the last
years systematically and successfully addressed the main technical
challenges of the accelerator project. Similarly, detailed detector
and physics studies confirm the ability to perform high-precision
measurements at CLIC. Recent preliminary implementation
studies show that a coherent staged implementation can be done,
leading to an impressive long-term and timely physics programme at the
energy frontier, beyond the LHC programme. New results at the LHC have
started to open the experimental door to some of the key physics
questions ahead of us where a future CLIC project can play a
determining role, and further LHC data at 8 and 14~TeV will lay out
the landscape in much more detail by 2016--2017.

With relevance to the European Strategy process three clear messages stand out for the next steps of the CLIC project:\vspace{2mm}
\begin{itemize}
\item A focused technical development programme on accelerators and
  detectors is needed in the
 period 2012--2016 to lay the ground work for a complete Project
 Implementation Plan for the CLIC project, to be ready by 2016. This
 will also entail a possible revision of the energy stages currently considered,
 taking into account LHC results at 8 and 14~TeV, and include
 a re-optimisation of the initial stages. Such a programme is well underway but 
 relies on continued and, in some cases, extended support by the CLIC collaborating 
 institutes and funding agencies throughout the full period.
\item A comprehensive common high-energy physics and detector study
programme is needed on the same time scale to assess the various options for CERN's future.
These include LHC luminosity upgrades (providing the yardstick for
comparisons), LC options, LHC energy upgrades, and other potential
facilities. Evaluation of the physics performance and capabilities of
each is necessary to decide on the future energy-frontier facility at
CERN after the LHC.
\item Sufficient funding and resources should be foreseen in the years
 {2017--2022} for advancing the project(s) chosen. Such support should be given a 
 common high priority within CERN, by the collaborating institutes and
 funding agencies, and within the European Commission programmes 
 that are relevant for developing projects of this type. 
 This requires coordinated resource planning between the LHC
 luminosity upgrades
 and the preparation of a future facility
 to be ready around 2030.  
\end{itemize}

\newpage
\bibliography{cdrbibliography.bib}

\end{document}